# Parallelism in Object Detection Deployment

OpenMP behavior in low resource and high stress mobile environment

Kaijun Zhang[†]
Computer Science
North Carolina State University
Raleigh, NC, US
kzhang29@ncsu.edu

## ABSTRACT

The state-of-the-art object detection models today all boasts incredible benchmarking performances, some toppling over 1000fps, which are all achieved under ideal environments and an equally impressive GPU. These numbers are perhaps not very reflective of day-to-day usage case. In most personal Android mobile devices, computational power, memory, and battery are all limited and need to be penny and dimed to inch out an acceptable Object Detection performance.

In this paper, I explored incorporating parallelism in the post-processing steps of object detection deployment using OpenMP, utilizing the device's shared memory, and compared its results from a variety of configurations. The configurations I hypothesized to be relevant were the number of threads, CPU affinity, and chunk size. The device this experiment is ran on is the Redmi Note 10 Pro, which has an ARM Cortex-A76 four core CPU.

The results of this experiment showed a maximum 2.3x speedup with four threads and dynamic work allocation compared to sequential post-processing. However, this configuration causes the overall inference time to increase by 2.7x, so the best overall configuration ended up being two threads with dynamic work allocation, which resulted in 1.8x speedup in post-processing and 2% speedup in the overall program.

## 1 Introduction

Object detection has become an essential component in various computer vision applications, including video surveillance, autonomous vehicles, robotics, and augmented reality. The rapid growth of deep learning techniques has significantly improved the accuracy and performance of object detection models. However, the high computational demands of these models pose challenges for deployment in low-resource and high-stress environments. In such settings, achieving real-time performance with limited computational resources is crucial.

Parallelism techniques have emerged as a promising solution to address the computational challenges associated with object detection deployment. One such technique is OpenMP, an open-source, portable, and scalable multi-platform shared-memory parallel programming model. OpenMP enables developers to parallelize their applications efficiently by providing a set of compiler directives, library routines, and environment variables. This parallelism allows for better utilization of multicore processors, which are common in low-resource environments.

Despite the potential benefits, the behavior of OpenMP in low-resource and high-stress environments has not been extensively studied. Understanding how OpenMP performs under these conditions is crucial to optimize object detection deployment and ensure reliable and efficient operation. This study aims to fill this gap by investigating the following research questions:

- How does OpenMP behave in low-resource and high-stress environments when applied to object detection deployment?
- What are the performance trade-offs and limitations associated with using OpenMP for parallelism in object detection tasks in such environments?
- How can OpenMP be optimized to improve the performance of object detection deployment in low-resource and high-stress settings?

To answer these questions, I conduct a series of experiments using state-of-the-art object detection model, Yolov6n, and benchmark them on various low-resource and high-stress scenarios. I analyze the performance, scalability, and efficiency of OpenMP under different configurations and provide guidelines for optimizing OpenMP-based parallelism in object detection deployment.

This paper is organized as follows: Section 2 reviews related work on object detection, parallelism techniques, and OpenMP. Section 3 describes the methodology, including the object detection models, experimental setup, and performance metrics. Section 4 presents the results and analysis of the experiments. Section 5 discusses the implications and limitations of the findings, and Section 6 concludes the paper with future research directions.

## 2  Related Works

This section provides an overview of the related work in parallelism in mobile devices, highlighting the existing research and the gaps this study aims to address.

### 2.1 Parallelism in Mobile Devices

Unlike stationary computing devices, energy source is not infinite on mobile devices and is a major constrain, which is why most current works that analyzes parallelism in the mobile platform focuses on the battery consumption of the parallel algorithm at different problem sizes or the number of threads [1].

There is also existing research that proposes dynamic core management, disabling less utilized cores during runtime to reduce power consumption [2]. However, there is limited research on the behavior and performance of OpenMP in object detection deployment, particularly on mobile Android devices where there is low resource and high computational demand.

This study aims to address this gap by investigating the performance of OpenMP-based parallelism in object detection deployment under various low-resource and high-stress scenarios. The findings of this study will contribute to the understanding of OpenMP behavior in such environments and provide guidelines for optimizing OpenMP-based parallelism for object detection tasks.

## 3  Methodology

In this section, I will describe the experimental setup, the object detection model and framework I used, the target device, and the specific problem I addressed. I will also outline the various configurations I tested to evaluate the performance of OpenMP-based parallelism in object detection deployment.

### 3.1  Experimental Set-up

I used the YOLOv6 Nano object detection model for the experiments, as it is a state-of-the-art model optimized for efficient mobile deployment [3]. For the deployment framework, I chose NCNN because it is specifically designed for low-resource and high-stress environments. The source code for this deployment is based on FeiGeChuanShu Android implementation [4].

The target device for the experiments was the Xiaomi Note 10 Pro, a budget smartphone with a Cortex-A76 CPU. Although this device has limited resources for running machine learning inferences, it provides a suitable platform for investigating the behavior of OpenMP in a low-resource and high-stress environment.

### 3.2  Problem Description

I focused on parallelizing the generate_yolox_proposals function in the post-processing stage of object detection. This function takes a feature map generated by the model, loops through the feature map, and creates "proposals" for objects that meet a minimum confidence threshold. I used OpenMP to spawn multiple threads to process the grid boxes in parallel (Figure 1).

```
1  PROCEDURE generate_yolox_proposals(grid_strides, feat_blob, prob_threshold, proposed_objects)
2      num_anchors ← grid_strides.size
3      num_class ← feat_blob.width - 5
4      #pragma parallel for
5      for i in num_anchors
6          #calculate box location and size
7          for j in num_class
8              #calculate box probability
9              if box_prob > 0.5
10                 Add box to proposed_objects
```

**Figure 1:** Pseudo code of generate_yolox_proposals. Num_anchors equals to 8400, num_class equal to 80.

### 3.3  Configurations

To assess the performance of OpenMP-based parallelism, I tested various configurations of the following parameters:

- Thread size: I varied the number of threads used in the OpenMP implementation, ranging from 1 (sequential) to 32 (eight times the core count of the target device).
- Scheduling strategies: I compared static and dynamic work allocation in OpenMP to evaluate the impact of different scheduling strategies on performance.
- Chunk size: I analyzed the effect of different chunk sizes on performance when using static and dynamic scheduling strategies.
- CPU core affinity: I examined the impact of different CPU core affinity configurations (spread, close,

and master) on performance when using static scheduling and a chunk size of 124.

For each configuration, I measured the time taken to complete the generate_yolox_proposals function, the total program runtime (time taken to process one image from beginning to end), and the CPU usage. I also profiled the CPU usage to identify any potential bottlenecks or resource contention.

The methodology enables us to systematically investigate the performance of OpenMP-based parallelism in object detection deployment under various low-resource and high-stress scenarios. The results of the experiments, presented in the next section, provide valuable insights into the behavior of OpenMP in such settings and inform future research on optimizing parallelism for object detection tasks.

# 4  Results

In this section, I present the results of the experiments on the performance of OpenMP-based parallelism in object detection deployment under various low-resource and high-stress scenarios. I will investigate the impact of different configurations, such as thread size, scheduling strategies, chunk size, and CPU core affinity, on the performance of the generate_yolox_proposals function in the post-processing stage.

## 4.1  Thread size

This experiment shows that using two threads with OpenMP leads to a nearly 2x speedup in the post-processing stage compared to the sequential approach. However, when the thread size exceeds double the core count, the latency spikes, and the process becomes unstable due to increased contention for CPU time and increased overhead for scheduling (Figure 2), which is a common rule-of-thumb behavior. However, I observed similar behavior with thread sizes of 4 and 8, which are at core size or double the core size (the common optimal setting), suggesting that the performance of post-processing could be influenced by other factors.

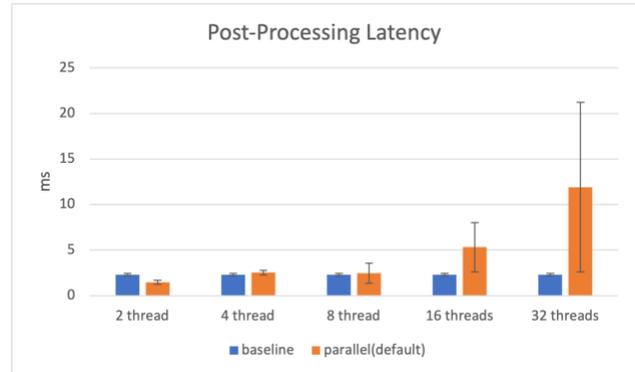

**Figure 2: Post-Processing Latency between different thread sizes**

## 4.2  Scheduling Strategy

Next, I compare the performance of static and dynamic work allocation in OpenMP. The results indicate that dynamic work allocation improves performance when thread sizes are low, but this improvement is offset by higher scheduling overhead at larger thread sizes (Figure 3). This suggests that dynamic work allocation may be beneficial for lower thread counts but not for the higher ones.

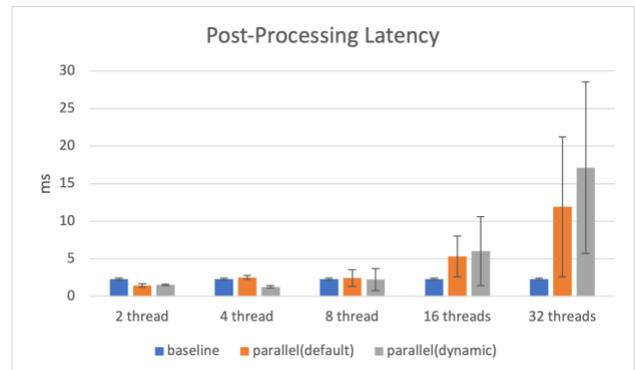

**Figure 3: Post-Processing Latency between different thread sizes and scheduling strategy**

## 4.3  Chunk Size

I also analyze the impact of varying chunk sizes on performance with both static and dynamic scheduling strategies. In both cases, I observed a concave trend, with an optimal chunk size leading to the best performance in the middle (Figure 4). This behavior is expected since there isn't much parallelization at higher chunk sizes, and low chunk sizes leads to more scheduling overhead. In general, both static and dynamic settings achieved similar results albeit static is slightly better at higher chunk sizes. This highlights the importance of finding the right balance between workload distribution and scheduling overhead.

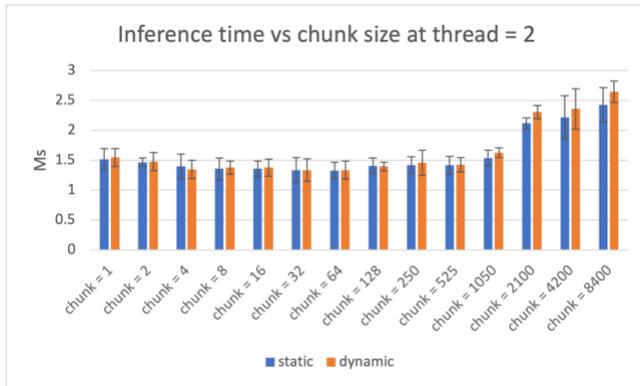

**Figure 4: Inference time at varying chunk sizes from 1 to number of grid strides (8400)**

### 4.4 CPU Core Affinity

The experiments on CPU core affinity reveal that there is not much variation between spread, close, and master configurations when using static scheduling and a chunk size of 124 (Figure 5). However, the master configuration appears to be slightly more stable at higher thread counts, due to reduced CPU contention in the other cores. It's also not surprising the results are very similar since there are only four cores for OpenMP to work with (whether to spread it or to keep it close). This result could also indicate that the operating system can make slightly better scheduling decision than OpenMP.

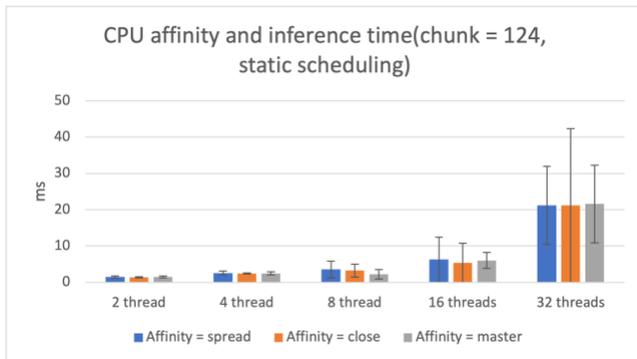

**Figure 5: CPU affinity and inference time at chunk size = 124 with static scheduling**

### 4.5 Total Application Runtime

Despite the observed improvements in the post-processing stage with certain configurations, the total program runtime, i.e., the time it takes to process one image from beginning to end, does not significantly improve. In fact, the total inference time almost triples when using four threads, even though the post-processing is fastest in this case (Figure 6). This is because NCNN, the framework used for the experiments, already employs multi-threading to speed up machine learning inference in the background, leading to high CPU contention, which causes more context switching and cache misses.

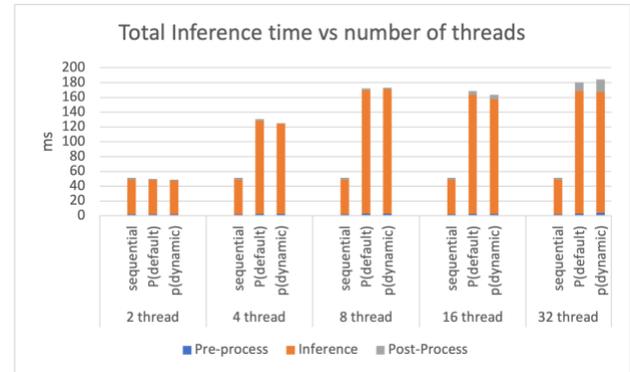

**Figure 6: Total inference time at varying number of threads (2 <= n <= 32).**

### 4.6 Profiling CPU usage

Profiling the CPU usage during the experiments shows that the CPU is fully occupied, with OpenMP creating multiple threads as expected (Figure 7). This further demonstrates the challenges of applying parallelism in low-resource and high-stress environments where the available resources are already heavily utilized.

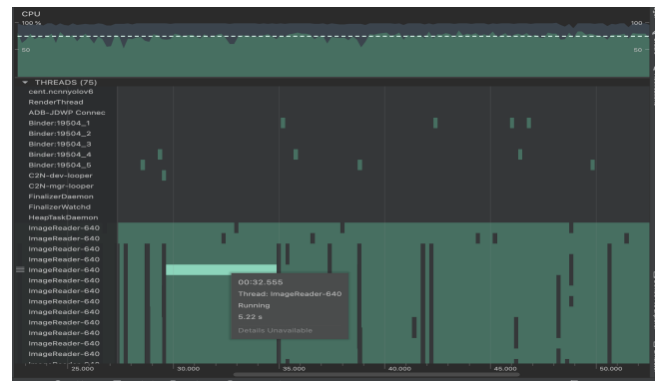

**Figure 7: CPU profiling in Android Studio. ImageReader threads are responsible for Pre-processing, Inference, and Post-processing**

In conclusion, the experiments show that it is possible to achieve a 2x speedup in the post-processing stage using OpenMP-based parallelism. However, this improvement only translates to a 2% optimization in the total program runtime, highlighting the challenges of deploying object detection models in low-resource and high-stress environments. The findings contribute to a better understanding of OpenMP behavior in such settings and provide a foundation for future research on optimizing parallelism for object detection tasks.

## 5  Implications and Limitations

In this section, I will discuss the implications of the findings for the deployment of object detection models in low-resource and high-stress environments, as well as the limitations of this study.

### 5.1  Implications

The results demonstrate that OpenMP-based parallelism can provide significant speedup in the post-processing stage of object detection deployment. This finding has important implications for the development and optimization of object detection models, as it suggests that parallelism can be leveraged to improve performance even in resource-constrained settings.

Moreover, the study highlights the importance of carefully selecting and tuning OpenMP configurations to achieve optimal performance. This insight can inform the development of optimization strategies and best practices for deploying object detection models in low-resource and high-stress environments.

Furthermore, the findings contribute to a better understanding of the behavior of OpenMP under non-ideal conditions, which is valuable for researchers working on mobile optimization of object detection deployment.

### 5.2  Limitations

While the study provides valuable insights into the performance of OpenMP-based parallelism in object detection deployment, it also has some limitations:

- Single device: The experiments were conducted on a single budget smartphone (Xiaomi Note 10 Pro). The performance of OpenMP may vary on other devices with different hardware configurations, making it important to validate the findings across a wider range of devices.
- Single model and framework: I used the YOLOv6 Nano object detection model and the NCNN framework for the experiments. The behavior of OpenMP may differ when using other object detection models or deployment frameworks.
- Focus on post-processing stage: this study focused on parallelizing the generate_yolo_proposals function in the post-processing stage. Other parts of the object detection pipeline, such as pre-processing and inference, may exhibit different behavior when parallelized using OpenMP.
- Limited configurations: Although I tested a variety of configurations, there may be other factors that impact the performance of OpenMP-based parallelism in object detection deployment, such as enabling OMP_DYNAMIC, which allows OpenMP to dynamically adjust the number of threads based on work load.

## 6  Conclusion and Future Work

In this paper, I investigated the performance of OpenMP-based parallelism in object detection deployment under low-resource and high-stress environments, using the YOLOv6 Nano object detection model and the NCNN framework on a Xiaomi Note 10 Pro device. The results demonstrate that OpenMP can provide significant speedup in the post-processing stage, even in resource-constrained settings. I also found that selecting and tuning OpenMP configurations carefully is essential for achieving optimal performance. It is also critical to be aware of the application wide context when attempting to parallelize one specific section.

### 6.1  Conclusion

This study contributes to a better understanding of the behavior of OpenMP in non-ideal conditions and has important implications for the development and optimization of object detection models. As object detection models continue to improve and real-world applications become more widespread, optimizing the deployment of these models on mobile devices will become increasingly important. The findings in this paper provide valuable insights for researchers and practitioners working on mobile optimization of object detection deployment.

### 6.2  Future Work

Based on the current findings and the limitations of this study, I propose several avenues for future research:

- Generalizing these findings by evaluating the performance of OpenMP-based parallelism on a wider range of devices and hardware configurations.

- Investigating the behavior of OpenMP when using other object detection models and deployment frameworks to assess the generalizability of these results.
- Exploring parallelization strategies for other stages of the object detection pipeline, such as pre-processing and inference, to identify additional opportunities for performance improvement.
- Developing optimization techniques for selecting and tuning OpenMP configurations to maximize performance in low-resource and high-stress environments.
- Investigating the use of other parallelization techniques, such as MPI, to further optimize the performance of object detection deployment in resource-constrained settings.

By addressing these research directions, we can continue to advance the field of object detection deployment and contribute to the development of more efficient and effective models and optimization strategies for resource-constrained environments.

## ACKNOWLEDGMENTS

I would like to thank Dr.Jiajia Li for her guidance and support throughout the semester. Her feedbacks and lectures have been critical in shaping this analyses.

## REFERENCES

[1] Malta Lindgren, (2021). "Evaluation of battery usage and scalability when performing parallel applications on mobile devices."

[2] Hwang, Y.-S. and Chung, K.-S. (2013) "Dynamic power management technique for multicore based embedded mobile devices," *IEEE Transactions on Industrial Informatics*, 9(3), pp. 1601–1612. Available at: https://doi.org/10.1109/tii.2012.2232299.

[3] Li, C. *et al.* (2022) "*Yolov6: A single-stage Object Detection Framework for industrial applications*", *arXiv.org*. Available at: https://arxiv.org/abs/2209.02976

[4] FeiGeChuanShu (2022) *Feigechuanshu/NCNN-Android-yolov6*, *GitHub*. Available at: https://github.com/FeiGeChuanShu/ncnn-android-yolov6 (Accessed: May 2, 2023).